# Minimum Rate Sampling and Spectrum Blind Reconstruction in Random Equivalent Sampling

Yijiu Zhao, Li Wang, Houjun Wang, Changjian Liu

**Abstract** The random equivalent sampling (RES) is a well-known sampling technique that can be used to capture a high-speed repetitive waveform with low sampling rate. In this paper, the feasibility of spectrum-blind multiband signal reconstruction for data sampled from RES is investigated. We propose a RES sampling pattern and its corresponding mathematical model that guarantees well-conditioned reconstruction of multiband signal with unknown spectral support. We give the minimum number of RES acquisitions that hold overwhelming probability to successfully reconstruct original signal. We demonstrate that for signal with specific spectral occupation, the number of RES acquisitions and the minimum sampling rate could be approached. The signal reconstruction is studied in the framework of compressive sampling (CS) theory. The eigen-decomposition and minimum description length (MDL) criteria are adopted to adaptively estimate the dimension of signal, and the number of unknowns of reconstruction problem is reduced. Experimental results are reported to indicate that, for a spectrum-blind sparse multiband signal, the proposed reconstruction algorithm for RES is feasible and robust.

## 1 Introduction

Digital system holds the promise to process the underlying signal in the digital framework. Since many sources of information are of analog nature, digital signal processing inherently hinges upon sampling a continuous-time signal to obtain a discrete-time representation. Consequently, sampling theory is a bridge between analog world and digital world. The Shannon sampling theorem governs all aspects in digital signal processing since it has been developed, which makes a mathematical statement that, to perfectly reconstruct a signal, the sampling rate is

Y. J. Zhao (Corresponding Author)
School of Automation Engineering, University of Electronic Science and Technology of China, Chengdu, China
e-mail: yijiuzhao@uestc.edu.cn

required to be no less than two times of the signal's maximum frequency [13]. As the signal frequency becomes higher (usually exceeds gigahertz (GHz)), the Shannon sampling theorem puts forward a great challenge to the analog-to-digital converter (ADC), and the improvement in state-of-the-art ADC is not sufficiently fast to catch up with the emerging applications in related fields. The conventional uniform sampling with a single high-speed ADC solution is impractical.

To address the challenge to ADC, many alternative sampling techniques have been developed. Time interleaved sampling is a widely used sampling approach in communication and radar system [6], [11], [18], and it provides an architectural possibility of improving the total sampling rate of system using multiple low-speed ADCs. Time interleaved sampling is a special class of multi-coset sampling or periodic non-uniform sampling. Specifically, time interleaved sampling could be cost effective or even the only possible solution in sampling high-speed transit signal due to the high cost or unavailability of a single ADC solution.

In the applications of data acquisition system, i.e. digital storage oscilloscope (DSO) [12], [19], the random equivalent sampling (RES) technique is a low cost and effective solution to capture signal with high frequency that is repetitive or could be repeatedly excited. Leveraging the in-synchrony between the sampling clock and the excitation signal, each time a sampling sequence is taken at a different position of a cycle of the analog waveform. In RES, after multiple acquisitions, a waveform with high equivalent sampling frequency can be composited from the samples taken by RES circuitry with a single low-speed ADC. Sampling efficiency and signal reconstruction accuracy hinge upon the randomness of time positions, and it could be realized by dithering the clock phase [14].

In practical implementation of RES, the relative sampling positions within a period are non-uniformly distributed. In the stage of reconstruction using time-alignment technique, multiple RES samples may be mapped to one sample, and only a few valid sample sequences could be obtained after a great amount of acquisitions. In order to collect enough valid sample sequences to reconstruct signal with desired accuracy, it may take significantly longer time to perform RES. Considering the inefficiency introduced by uneven distribution of time positions, in this work, we present an adaptive spectrum sensing algorithm in the framework of compressive sampling (CS) to improve RES.

CS [3], [4] is a new signal processing theory, which can reconstruct a high-speed sparse signal from sub-Nyquist samples. In our previous work [20], [22], we have demonstrated the feasibility and efficiency of incorporating CS with RES, and the proposed CS-based signal reconstruction can significantly improve RES performance. However, CS-RES is only applicable for harmonic sparse signal and does not work for sparse multiband signal that is widely used in communication and radar system. Different from our previous work, in this paper, we model the RES in the frequency domain. Based on the RES model, the lower bound of the minimum sampling rate is given, and the minimum number of RES acquisitions is suggested. In the framework of CS, an adaptive spectrum sensing algorithm is proposed for multiband signal that is with unknown active band occupation and unknown number of active bands. The band support can be estimated from the RES samples using the proposed spectrum sensing algorithm. According to the estimated band support, the signal can be reconstructed from the corresponding columns of the measurement matrix.

The rest of this paper is structured as follows: The formulation of the problem is introduced in Section 2. A brief review of random equivalent sampling and minimum sampling rate theorem is present and stated in Section 3. Section 4 introduces a CS-based adaptive blind spectrum multiband signal reconstruction algorithm for RES samples. In section 5, experimental results are reported to evaluate the proposed algorithm. Conclusion is summarized in section 6.

**2 Formulation of the Problem**

Sparse multiband signal is the prime focus of our work. A sparse multiband signal is bandlimited, squared integrable, continuous-time signal, and its energy concentrated in one or more disjoint frequency bands. Considering a sparse multiband signal $x(t)$, and its Fourier transform is

$$X(f) = \int_{-\infty}^{+\infty} x(t) e^{-j2\pi ft} dt . \qquad (1)$$

Define $\mathcal{B}(\mathcal{F})$ as the set containing all multiband signal $x(t)$, it satisfies the following:
(1) $\mathcal{F} \subset [-1/2T, 1/2T]$ ($1/T$ is the Nyquist frequency of underlying signal), and $X(f) = 0$ if $f \notin \mathcal{F}$.
(2) $\mathcal{F} = \cup_{i=1}^{K} \mathcal{F}_i$, $X(f)$ consists of $K$ union of frequency intervals, the number of frequency interval is unknown, and $\mathcal{F}_i$ does not exceed $B$ ($B$ is the maximum band width).

A typical spectrum-sparse multiband signal is shown in Fig. 1. Let $f_{min}$ and $f_{max}$ be the minimum frequency and the maximum frequency respectively. Let $[\mathcal{F}]$ denote the spectral span, which is the smallest interval containing $\mathcal{F}$, and $[\mathcal{F}] = [f_{min}, f_{max}]$. Then the spectral occupation ratio $\Omega$ satisfying $\Omega \geq \lambda(\mathcal{F})/|[\mathcal{F}]|$ ($\lambda(\mathcal{F})$ is the Lebesgue measure of $\mathcal{F}$, and $0 < \Omega < 1$). Since any signal can be shifted to the origin by multiplication of the signal in the time domain by a signal with specific frequency, $f_{min}$ is set to zero, and $f_{max} = 1/2T$.

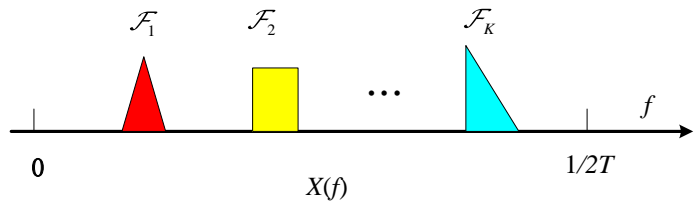

**Fig. 1** Spectrum-sparse multiband signal.

In this work, we wish to perfectly reconstruct $x(t) \in \mathcal{B}(\cdot)$ from a set of sample points. The problem we consider is to find a universal sampling model and an adaptive spectrum blind reconstruction algorithm. The universal sampling model should satisfy the stability condition and achieve the minimum sampling rate. On the other hand, in the reconstruction stage, there is no prior information of position of active bands of signal, and the number of active bands is also unknown in advance.

## 3 Signal Sub-Nyquist Sampling

3.1 Random Equivalent Sampling

In instrumentation application, RES provides an architectural solution to capture high-speed repeatedly excited signal using an ADC clocked at the sub-Nyquist rate. A waveform with high equivalent sampling could be composited from RES samples by adopting time-alignment method. Fig. 2 depicts the basic principle of RES. The underlying signal is repeatedly excited under control of the excitation pulse as shown in the solid line on the top row. In each acquisition, a sampling sequence is taken using an ADC clocked at $1/T_s$ that is much lower than the Nyquist rate. The relative sampling position of $m$th acquisition $\Delta t_m$ ($0 < \Delta t_m \leq T_s$) between the excitation pulse and the immediately sampling pulse is measured by a time-to-digital converter (TDC) circuitry [21]. In RES, it is assumed that the position between the underlying signal and the excitation pulse is fixed, and the excitation pulse provides a fixed reference point to store sampling sequence using time-alignment method.

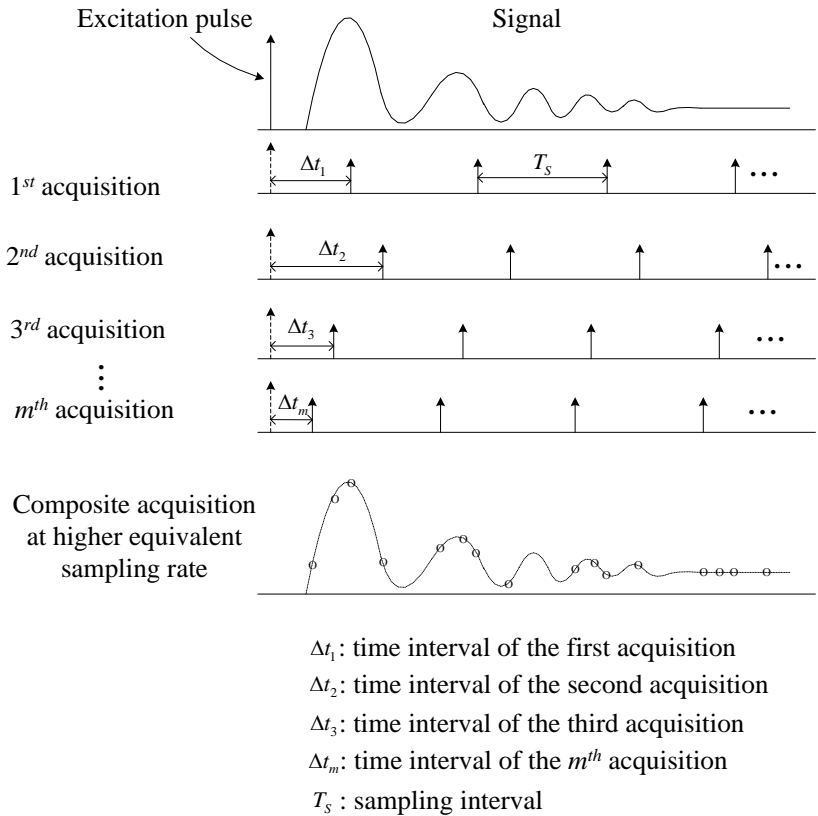

$\Delta t_1$: time interval of the first acquisition
$\Delta t_2$: time interval of the second acquisition
$\Delta t_3$: time interval of the third acquisition
$\Delta t_m$: time interval of the $m^{th}$ acquisition
$T_s$: sampling interval

**Fig. 2** Basic principle of RES, a waveform with high equivalent sampling rate is composited from sampling sequence after multiple random acquisitions.

In practical application of RES, sampling interval $T_s$ need to be partitioned into $P$ equally spaced bins of duration $T_e$, and $f_e = 1/T_e$ is also called the equivalent sampling frequency. In order to reconstruct signal without aliasing, $f_e$ should be no smaller than the signal's Nyquist sampling rate. For expression simplicity, hereafter, $T_e$ is referred to as $T$. In conventional time-alignment reconstruction method, the success of RES depends on the assumption that the time interval $\Delta t$ of each sampling sequence is evenly distributed over time interval $(0, T_s]$. In our previous work [20], we have comprehensively analyzed the inefficiency of RES introduced by the uneven distribution of the time interval. In this work, we remodel the RES behavior in the frequency domain and wish to propose an adaptive spectrum blind sensing algorithm.

Assume $x(n \cdot T)$ is the uniform Nyquist sampling sequence of $x(t)$, which is with the equivalent sampling rate $1/T$. According to the above analysis, $x(n \cdot T)$ satisfies the Shannon sampling theorem, and it contains all the information of $x(t)$. Let $M$ be the number of RES acquisitions, considering the $m$th RES acquisition ($1 \leq m \leq M$), the $m$th sampling sequence can be expressed as follow:

$$x_m[n] = x(n \cdot T_s + \Delta t_m) \\ = x(n \cdot P \cdot T + \Delta t_m), \quad n \in \mathbb{Z} \tag{2}$$

and its discrete-time Fourier transform (DTFT) [10] is

$$X_m\left(e^{j2\pi fPT}\right) = \sum_{n=-\infty}^{+\infty} x_m[n]\exp(-j2\pi fnPT) \\ = \sum_{n=-\infty}^{+\infty} x(nPT + \Delta t_m)\exp(-j2\pi fnPT). \tag{3}$$

Since $x_m[n]$ is the shifted and down sampled version of $x(nT)$, the direct calculations link the known DTFT $X_m\left(e^{j2\pi fPT}\right)$ of $x_m[n]$ to the unknown Fourier transform (FT) $X(f)$ of $x(t)$:

$$X_m\left(e^{j2\pi fPT}\right) = \sum_{n=-\infty}^{+\infty} x(nPT + \Delta t_m)\exp(-j2\pi fnPT) \\ = \frac{1}{PT} \sum_{r=-(P-1)/2}^{(P+1)/2} X\left(f - \frac{r}{PT}\right)\exp\left(j2\pi\Delta t_m\left(f - \frac{r}{PT}\right)\right). \tag{4}$$

Note that, for every $1 \leq m \leq M$, $f$ is restricted to interval

$$\mathcal{F}_0 = \left[-\frac{1}{2PT}, \frac{1}{2PT}\right). \tag{5}$$

Here $P$ is assumed to be odd value, for the case of even value, the range of $r$ and $\mathcal{F}_0$ should be redefined [9]. The restriction to $\mathcal{F}_0$ removes the dependence on $f$ in the summation limits since within this interval is a linear combination of a particular (finite) set of spectral segments of $x(t)$. Eqn. (4) ties the known random sampling sequence and the unknown signal $x(t)$, and it can be rewritten as the matrix-vector

formulation:

$$\begin{bmatrix} y_1(f) \\ y_2(f) \\ \vdots \\ y_M(f) \end{bmatrix} = \begin{bmatrix} \phi_{1,1} & \phi_{1,2} & \cdots & \phi_{1,P} \\ \phi_{2,1} & \phi_{2,1} & \cdots & \phi_{2,P} \\ \vdots & \vdots & \ddots & \vdots \\ \phi_{M,1} & \phi_{M,2} & \cdots & \phi_{M,P} \end{bmatrix} \begin{bmatrix} x_1(f) \\ x_2(f) \\ \vdots \\ x_P(f) \end{bmatrix}$$

$$\Leftrightarrow \mathbf{y}(f) = \mathbf{\Phi}\,\mathbf{x}(f) \quad \forall f \in \mathcal{F}_0 \tag{6}$$

where $\mathbf{y}(f)$ is a column vector of length $M$, $\mathbf{\Phi}$ is an $M \times P$ measurement matrix, and their elements are as following

$$y_m(f) = PT \cdot \exp(-j2\pi f \Delta t_m) X_m\left(e^{j2\pi fPT}\right) \tag{7a}$$

$$\phi_{m,n} = \exp\left(-j\frac{2\pi}{P} \cdot \frac{\Delta t_m}{T} \cdot r_n\right) \tag{7b}$$

where $n = 1, 2, \ldots, P$, and $r_n = -(P+1)/2 + n$. $\mathbf{x}(f)$ can be treated as a vector that contains $P$ unknown frequency intervals by slicing $X(f)$. The relation between $\mathbf{x}(f)$ and $X(f)$ is depicted in Fig. 3. For each $f$, the $n$th element of $\mathbf{x}(f)$ is

$$x_n(f) = X\left(f - \frac{r_n}{PT}\right), \quad f \in \mathcal{F}_0. \tag{8}$$

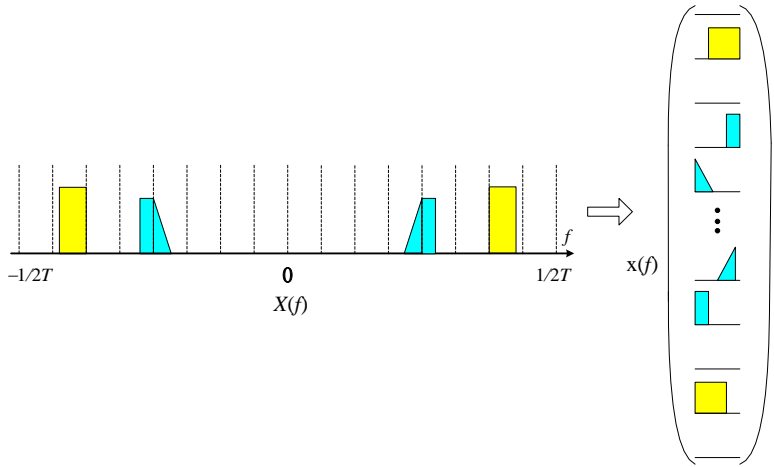

**Fig. 3** $X(f)$ is partitioned into $P$ equal width frequency intervals that are the elements of vector $\mathbf{x}(f)$.

3.2 Minimum Sampling Rate

In RES sampling, the signal is sampled using an ADC clocked at the sub-

Nyquist sampling rate $1/T_s$. If the conventional time-alignment method is adopted to reconstruct signal, the sampling rate could be arbitrarily low. However, the bins number $P$ is proportional to $T_s$. For the larger $P$, more sampling sequences should be captured, and a longer time needs to be taken. To make the matter even worse, the uneven distribution of time interval $\Delta t$ may make it impossible to capture enough valid sampling sequences that are needed for the reconstruction with desired accuracy.

In order to avoid the inefficency introduced by the uneven distribution, the model (6) is adopted. Considering the multiband signal $x(t) \in \mathcal{B}(\cdot)$, the minimum sampling rate of RES cannot be arbitrarily low.

*Theorem* 1 (*minimum sampling rate*): Let $x(t) \in \mathcal{B}(\cdot)$ be a multiband signal, for RES sampling, its minimum sampling rate is $1/T_s = B$.

*Proof*: For RES sampling, $T_s = PT$. According to (8), $1/T$ is sliced into $P$ equal width spectral bins. If the multiband signal $x(t) \in \mathcal{B}(\cdot)$, it contains multiple disjoint active bands with maximum bandwidth of $B$. Therefore, $f \in \mathcal{F}_k$, we have expression as following

$$f + \frac{i}{PT} \notin \mathcal{F}_k \quad \forall i \neq 0 \tag{9}$$

and it implies

$$f + \left|\frac{i}{PT}\right| \geq f + B \quad i \in \mathbb{Z}, \forall i \neq 0 \tag{10}$$

and take $i = 1$

$$\frac{1}{PT} \geq B. \tag{11}$$

Since $T_s = PT$, $1/T_s \geq B$, and the proof is end.

As we know, for a signal with known spectral support, the Landau rate [7] is $KB$. However, the Landau theorem does not apply to signal $x(t) \in \mathcal{B}(\cdot)$. Let $f_{total}$ denote the total sampling rate of RES, for a reconstruction with $M$ acquisitions, it can be expressed as

$$f_{total} = \frac{M}{PT}. \tag{12}$$

*Theorem* 2 (*minimum total sampling rate /minimum number of RES acquisitions*): For $x(t) \in \mathcal{B}(\cdot)$, the minumum total sampling rate is

$$f_{total} = \min\left\{2KB, \frac{1}{T}\right\} \tag{13}$$

and the minimum number of RES acquisitions is

$$M = \min\{2KPBT, P\}. \tag{14}$$

The proof of theorem 1 in [9] can be adopted to show that the minimum total sampling rate satisfies (13). Combining (12) and (13), (14) can be proved. Obviously, the RES could achieve sub-Nyquist sampling if the spectral occupation ratio $\Omega < 0.5$. And for $\Omega \geq 0.5$, the equivalent sampling rate of RES is required to be no less than the Nyquist rate.

**4 CS-based Adaptive Spectrum-Blind Reconstruction**

In this work, we wish to reconstruct signal in the framework of CS theory, and the underlying signal is with the spectral occupation ratio $\Omega < 0.5$ (less than 50% of band range is active). Eqn. (6) models the relation between the known RES samples $\mathbf{y}(f)$ and the unknown original signal $\mathbf{x}(f)$. However, in order to achieve the sub-Nyquist sampling, $M \ll P$, and (6) is an underdetermined problem. Fortunately, when incorporating the structure of $\mathbf{x}(f)$ as prior, the number of unknown could be reduced, and we can find a solution to (6).

Consider a class $\mathcal{M} \subset L_2(\mathbb{R})$ of signals, for every signal $\mathbf{x}_1 \in \mathcal{M}$, $\mathbf{x}_2 \in \mathcal{M}$, a set of points $\Psi$ is called a sampling set for $\mathcal{M}$, if it satisfies the stability condition [1] as following

$$\alpha\|\mathbf{x}_1 - \mathbf{x}_2\|_{\mathcal{M}} \leq \|\mathbf{x}_1(\Psi) - \mathbf{x}_2(\Psi)\|_{l_2} \leq \beta\|\mathbf{x}_1 - \mathbf{x}_2\|_{\mathcal{M}} \tag{15}$$

where $0 < \alpha \leq \beta < \infty$, $\mathbf{x}(\Psi)$ is the sampling sequence of $\mathbf{x}$ at position $\Psi$. The stability condition (15) guarantees not only unique solution but also a small reconstruction error. In RES sampling, the sampling pattern $\Psi$ is random, and a random sampling pattern is stable with high probability [2].

The problem we considered is to develop an adaptive reconstruction algorithm for signal with unknown spectral occupation and unknown number of active bands. To reconstruct signal $x(t)$, we must first determine the support of $\mathbf{x}(f)$ from $\mathbf{y}(f)$. Let $S$ denote the index set that marks the locations of the nonzero elements of $\mathbf{x}(f)$, its expression can be written as

$$S = \bigcup_{f \in \mathcal{F}_0} I(\mathbf{x}(f)). \tag{16}$$

Once $S$ is obtained, perfect reconstruction of $\mathbf{x}(f)$ is possible for every $f \in \mathcal{F}_0$, and reconstruction problem (6) can be rewritten as:

$$\mathbf{y}(f) = \mathbf{\Phi}_S \mathbf{x}^S(f) \tag{17}$$

where matrix $\mathbf{\Phi}_S \in \mathbb{C}^{P \times |S|}$ is constructed from the columns of $\mathbf{\Phi}$ indexed by $S$, and $\mathbf{x}^S(f)$ is the elements of $\mathbf{x}(f)$ indexed by $S$. If the index set of $\mathbf{x}(f)$ satisfies $|S| \leq \text{rank}(\mathbf{\Phi})$, then, the matrix $\mathbf{\Phi}_S$ is full column rank. Eqn. (17) is a determined or an over determined linear system, and it can be inverted using a pseudoinverse.

$$\begin{aligned}\mathbf{x}^S(f) &= \mathbf{\Phi}_S^\dagger \mathbf{y}(f) \\ &= \left([\mathbf{\Phi}_S]^H \mathbf{\Phi}_S\right)^{-1} [\mathbf{\Phi}_S]^H \mathbf{y}(f)\end{aligned} \quad (18)$$

where "$H$" and "$\dagger$" denote Hermitian transpose and pseudoinverse respectively.

Signal reconstruction based on model (6) could be divided into two stages: support estimation in the frequency domain and signal reconstruction in the time domain. Our focus in this work is to estimate the spectrum of signal.

Support recovery refers to the process of identifying which elements of $\mathbf{x}(f)$ contain the active bands that comprise $x(t)$. Since each element of $\mathbf{x}(f)$ is a spectral bin of $X(f)$ with band width of $1/(PT)$ Hz, identifying the nonzero spectral bins can only determine the true band support with a resolution of $1/(PT)$ Hz. Support can be teased out using an eigen-decomposition of the covariance matrix of the channel outputs and in particular on the covariance matrix's range space [5].

Considering the $M \times M$ covariance matrix of $\mathbf{y}(f)$,

$$\begin{aligned}\mathbf{R}_y &= \int_{-1/(2PT)}^{1/(2PT)} \mathbf{y}(f)\mathbf{y}(f)^H \, df \\ &= \mathbf{\Phi}\left[\int_{-1/(2PT)}^{1/(2PT)} \mathbf{x}(f)\mathbf{x}(f)^H \, df\right]\mathbf{\Phi}^H \\ &= \mathbf{\Phi}\mathbf{R}_x\mathbf{\Phi}^H\end{aligned} \quad (19)$$

where $\mathbf{R}_x$ is the covariance matrix of the spectral bins of $\mathbf{x}(f)$, and $\mathbf{\Phi}\mathbf{\Phi}^H = \mathbf{I}$. Since $\mathbf{x}(f)$ is $|S|$-sparse, rank($\mathbf{R}_x$) = $|S|$, and $\mathbf{R}_y$ can be reduced as $\mathbf{\Phi}_S[\mathbf{R}_x]_S[\mathbf{\Phi}_S]^H$. Because RES satisfies the stability condition (15), rank($\mathbf{\Phi}_S$) = $|S|$, and it implies:

$$\text{rank}(\mathbf{R}_y) = |S|. \quad (20)$$

By employing the eigen-decomposition of $\mathbf{R}_y$, the space of $\mathbf{R}_y$ could be divided into range space and null space. Define $\Sigma_r$ and $\Sigma_n$ denote the diagonal matrices containing the nonzero and zero eigenvalues respectively, the eigen-decomposition has the following expression

$$\begin{aligned}\mathbf{R}_y &= \mathbf{U}_r \Sigma_r [\mathbf{U}_r]^H + \mathbf{U}_n \Sigma_n [\mathbf{U}_n]^H \\ &= \mathbf{U}_r \Sigma_r [\mathbf{U}_r]^H\end{aligned} \quad (21)$$

where $\mathbf{U}_r$ and $\mathbf{U}_n$ are eigenvectors matrices. According to (19) and (21), we have $I(\mathbf{R}_y) = I(\mathbf{U}_r)$, and $I(\mathbf{U}_r) = I(\mathbf{\Phi}_S)$. Therefore, the columns of $\mathbf{\Phi}$ indexed by $S$ should lie in the space spanned by the columns of $\mathbf{U}_r$, and it can be determined by adopting the orthogonal matching pursuit (OMP) algorithm [15]. Once the estimation of $S$ is obtained, the original signal in the time domain can be reconstructed [8].

In previous works of spectrum-blind signal reconstruction algorithms, either the number of active bands is assumed to be known or a hard threshold is adopted [16]. However, in practical application, the number of active bands of underlying signal may not be known. On the other hand, due to the noise introduced by sampling stage, it is hard to set a suitable threshold. The index set can be identified by the large eigenvalues. In this work, we determine the number of active bands using the information theoretic criteria: minimum description length (MDL) [17].

Define $\lambda_i$ ($1 \leq i \leq M$) and $d$ denote the estimated eigenvalue of $\mathbf{R}_y$ and dimension of range space of signal respectively. Then, the MDL criteria of signal is given by

$$\text{MDL}(d) = -\lg \left( \frac{\prod_{i=d+1}^{M} \lambda_i^{1/(M-d)}}{\frac{1}{M-d} \sum_{i=d+1}^{M} \lambda_i} \right)^{(M-d)L} + \frac{1}{2} d(2M-d) \lg L \qquad (22)$$

where $\lambda_1 \geq \ldots \geq \lambda_M$, and $L$ is the length of the sampling sequence. The MDL yields a consistent estimate. In large-sample limit, MDL($d$) is minimized for $d = |S|$. By employing the MDL criteria, the dimension of range space of underlying signal can be adaptively determined.

## 5 Experimental Results

In this section, the performance of the proposed spectrum-blind signal reconstruction algorithm for RES is evaluated. The potential application of the proposed algorithm is the sparse multiband signal with unknown band locations and unknown number of bands. In all experiments, the underlying signal is shown in the following:

$$x(t) = \sum_{i=1}^{N} \sqrt{E_i B} \cdot \text{sinc}(B(t-\tau_i)) \cos(2\pi f_i (t-\tau_i)) \qquad (23)$$

where $N$ is the number of active bands, $E_i$ is the energy coefficient, $B$ is the band width, $\tau_i$ is the time offsets, and $f_i$ is the carrier frequency. In the reconstruction stage, all parameters but $B$ will be unknown. The signal-to-noise ratio (SNR) is used as a metric to evaluate the proposed algorithm, and it is defined as:

$$\text{SNR(dB)} = 20 \cdot \log_{10} \left( \frac{\|X(f)\|}{\|X(f) - X^*(f)\|} \right) \qquad (24)$$

where $X^*(f)$ and $X(f)$ are the reconstructed signal and the original signal respectively.

In all following experiments, the equivalent sampling rate $1/T = 1$ GHz, the RES sampling rate $1/T_s = 10$ MHz, $P = T_s/T = 100$.

The first experiment tests the feasibility of the proposed spectrum-blind reconstruction algorithm. The underlying signal is noise-free. We set the experiment parameters to $N = 2$, $E = \{3, 2\}$, $\tau = \{12, 15\}$ us, $f = \{115.3, 370.7\}$ MHz, and $B = 8$ MHz. $M = 12$ random RES acquisitions are performed, and sampling time interval $\Delta t$ is measured to construct the measurement matrix $\Phi$. Fig. 4(a) shows the comparison between the original signal and the reconstructed signal. Obviously, the signal could be perfectly reconstructed from RES samples using the proposed algorithm. Fig. 4(b) depicts the reconstructed spectrum of signal, and 4 active bands (the number of the active bands of complex-valued signal is $K = 2N$) are successfully identified. The diversity of index set is $S = \{11, 12, 36, 37, 62, 63, 88\}$. The MDL estimated dimension of range($\mathbf{R}_y$) is $d = 7$, and it is consistent with $|S|$. Although $B$ is smaller than the duration of spectral bin, the active band may be sliced into two adjacent bins. Therefore, $d$ satisfies $K \leq d \leq 2K$.

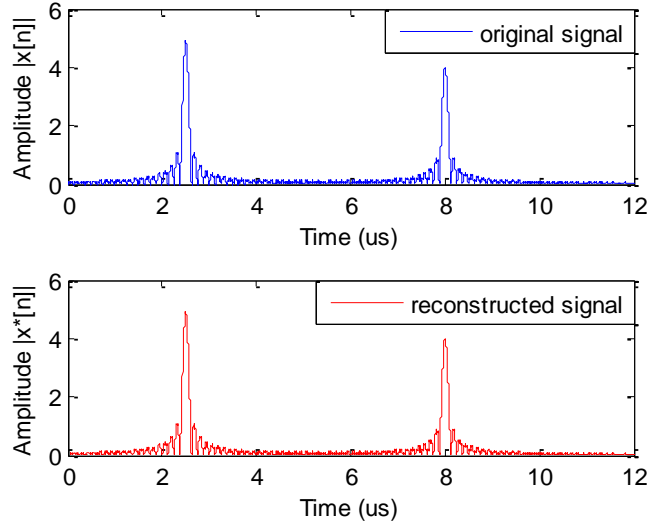

(a)

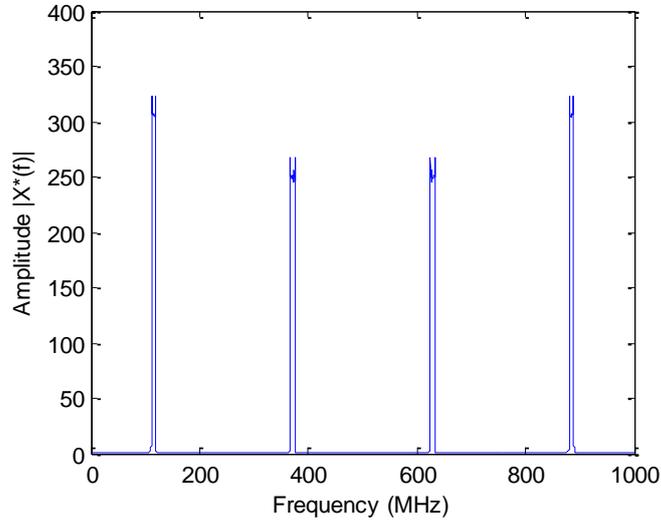

(b)

**Fig. 4** Spectrum-blind signal reconstruction. (a) Comparison between the original signal and reconstructed signal. (b) Spectrum of the reconstructed signal.

Next, the performance of reconstruction with respect to the numbers of RES acquisitions is investigated. For the range of 10 to 60 RES acquisitions in increment of 5 RES acquisitions, 200 random trials are performed for each specific value. In this experiment, two spectrum-blind signals with different sparsity level $N$ are considered. All parameters but $B$ and $f_i$ are randomly selected. $B$ is set to 8 MHz, and $f_i$ is selected to satisfy that the dimensions of index set $S$ of underlying signals are $d = 8$ and $d = 11$ respectively. The averaged SNRs of 200 trials at each acquisition number $M$ are plotted in Fig. 5. It is clear that the performance of reconstruction is improved with the increase of $M$ (equivalent to the total sampling rate of RES). For a specific value of $M$, the reconstruction of low dimensional signal exhibits higher SNR than that of high dimensional signal.

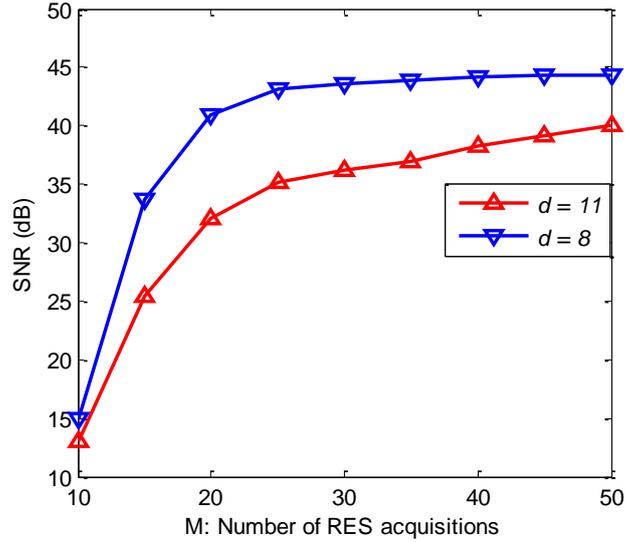

**Fig. 5** Reconstruction performance with respect to the number of RES acquisitions.

In practical application, the signal may be disturbed by noise, or the noise may be introduced in the sampling stage. In this experiment, we consider more practical situation that the underlying signal is corrupted by noise. In sampling stage, the Gaussian distributed noise is added in signal. The experiment signal parameters settings are the same as that of the first experiment. Noise levels over the range of 5 to 50 dB increment of 5 dB are evaluated, and 200 random trials are repeated for each specific noise level. The output SNRs are averaged and depicted in Fig. 6. Note that the proposed spectrum-blind signal algorithm is robust against additive Gaussian noise. For a specific input SNR, the reconstruction from more sampling sequences (larger $M$) yields higher SNR.

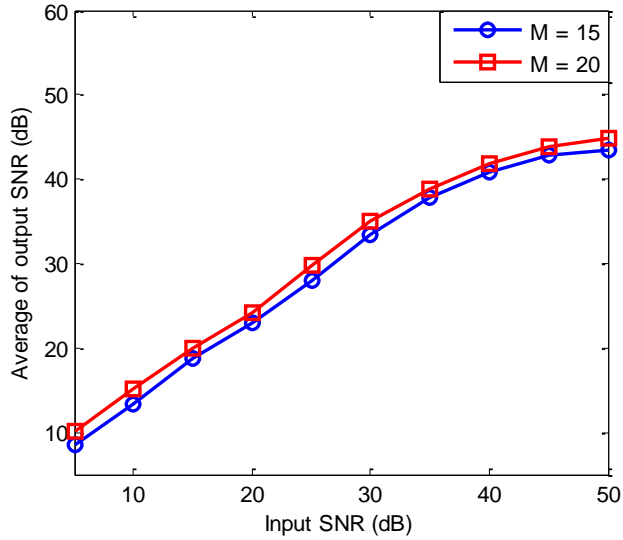

**Fig. 6** Reconstruction test for different noise levels.

## 6 Conclusion

In this paper, we propose a spectrum-blind signal reconstruction algorithm for RES in the framwork of CS theory, and the proposed algorithm can mitigate the influence caused by the uneven distribution of RES sampling time interval. This work indicates that the dimension of sparse multiband signal could be reduced by slicing the spectrum of signal. The mathmetical behavior of RES is modeled in the frequency domain, and the measurement matrix is constructed. When the spectral occupation rate is lower than 0.5, RES can achieve minimum total sampling rate, and the number of RES acquisitions also can be minimized. The active spectral bins are identified using the eigen-decomposition technique. Combining the MDL criteria, the dimension of index set could be determined. Experiments demonstrate that the proposed algorithm is able to adaptively determine the dimension of signal, and it is feasible and robust against additive Gaussian noise. Spectrum-blind sparse multiband signal could be accurately reconstructed with overwhelming probability.

## Acknowledgements

This work is supported in part by the National Natural Science Foundation of China (Grant No. 61301264), in part by the Ph.D. Programs Foundation of Ministry of Education of China (Grant No. 20130185120019), and in part by the Fundamental Research Funds for the Central Universities (Grant No. ZYGX2013J089).